# HARMONIC MEAN AS A DETERMINANT OF THE GENETIC CODE

Miloje M. Rakočević

Department of Chemistry, Faculty of Science, University of Niš, Serbia
(E-mail: milemirkov@open.telekom.rs; www.rakocevcode.rs)

**Abstract.** It is shown that there is a sense in splitting Genetic Code Table (GCT) into three parts using the harmonic mean, calculated by the formula H (a, b) = 2ab / (a + b), where *a* = 63 and *b* = 31.5. Within these three parts, the amino acids (AAs) are positioned on the basis of the validity of the evident regularities of key parameters, such as polarity, hydrophobicity and enzyme-mediated amino acid classification. In addition, there are obvious balances of the number of atoms in the nucleotide triplets and corresponding amino acid groups and/or classes.

## 1. Introduction

In a previous work we showed (on the model of a binary tree, 000000-111111, that is 0 - 63) that the *golden mean* is a characteristic determinant of the genetic code (GC) (Rakočević, 1998), and in this paper we will show that the same is valid for *the harmonic mean*, if Genetic Code Table (GCT) is divided into three equal[1] parts and if the ordinal codon number is the same as in R. Swanson's concept (Swanson, 1984; Rakočević, 1998, Fig. 1, p. 284).

According to the concept of R. Swanson, the order follows this sequence through the ordinal numbers: YUN, YCN, RUN, RCN / YAN, YGN, RAN, RGN. [Perhaps Negadi's order (UUN, UCN, …, GAN, GGN) (Negadi, 2009), as a sub-variant, corresponds with this order, but that is a question for further researches.] However, as we can see from GCT, the other (chemical) possibility is: NUN, NCN / NAN, NGN in accordance with the ordinal numbers 0-15, 16-31, 32-47 and 48-63 within four columns. In this paper it will become evident that this second order does not allow a chemically adequate division into three harmonic parts. (Note: Y for pYrimidine, R for puRine and N for aNy of four bases.)

In fact, two concepts of R. Swanson are taken into account - Gray Code Model of the genetic code as the first, and the Codon Path Cube as the second one. Both models start with binary values of nucleotide bases: U = 00, C = 01, A = 10 and G = 11. In the first concept, R. Swanson showed that codon binary values (and the codon ordinal number) are derived from the binary values of bases, from zeroth codon UUU = 000000 to the last one GGG = 111111; and we have shown that the same records can be applied to the binary-code tree (Rakočević, 1998), as well as in the GCT as the ordinal number of the codons (Table 1 and Appendix A).

On the other hand, we have also shown (Rakočević, 1988) that on the Codon Path Cube a special kind of quantitative "weight" of codons can be calculated, following R. Swanson's idea that "the three edges of the cubes represent the three positions in a codon" (Swanson, 1984, p. 189). According to these "weights", calculated on the basis of the binary numeral system (q = 2), codons are arranged in four "floors" of GCT with the following ordinal numbers: 0-15 in the first column, 2-17 in the second, 4-19 in the third and 6-21 in the fourth column. For better understanding, we provide an example of calculation for the last codon, the codon GGG: $(3 \times 2^2) + (3 \times 2^1) + (3 \times 2^0) = 21$ (Rakočević, 1988). In this way, the number 21 appears to be a specific and important determinant of the genetic code one more time.

---

[1] The codon space is made of 64 codons with 63 intervals between them (Table 1). When we say "three equal parts" that means 63 : 3 = 21. In the first part there are 22 codons, from zeroth UUU to 21st, GUC. In the second part there are 21 codons, from 22nd, GUA to the 42nd, UGA. Finally, within the third part there are also 21 codons, from 43rd, UGG to the last GGG.

| 1st lett. | 2nd letter | | | | 3rd lett. |
|---|---|---|---|---|---|
| | U | C | A | G | |
| U | 00. UUU  **F**-II<br>01. UUC<br>02. UUA  **L**-I<br>03. UUG | 08. UCU<br>09. UCC  **S**-II<br>10. UCA<br>11. UCG | 32. UAU  **Y**-I<br>33. UAC<br>34. UAA  CT<br>35. UAG | 40. UGU  **C**-I<br>41. UGC<br>42. UGA  CT<br>43. UGG  **W**-I | U<br>C<br>A<br>G |
| C | 04. CUU<br>05. CUC  **L**-I<br>06. CUA<br>07. CUG | 12. CCU<br>13. CCC  **P**-II<br>14. CCA<br>15. CCG | 36. CAU  **H**-II<br>37. CAC<br>38. CAA  **Q**-I<br>39. CAG | 44. CGU<br>45. CGC  **R**-I<br>46. CGA<br>47. CGG | U<br>C<br>A<br>G |
| A | 16. AUU<br>17. AUC  **I**-I<br>18. AUA<br>19. AUG  **M**-I | 24. ACU<br>25. ACC  **T**-II<br>26. ACA<br>27. ACG | 48. AAU  **N**-II<br>49. AAC<br>50. AAA  **K**-II<br>51. AAG | 56. AGU  **S**-II<br>57. AGC<br>58. AGA  **R**-I<br>59. AGG | U<br>C<br>A<br>G |
| G | 20. GUU  **V**-I<br>21. GUC<br>22. GUA  **V**-I<br>23. GUG | 28. GCU<br>29. GCC  **A**-II<br>30. GCA<br>31. GCG | 52. GAU  **D**-II<br>53. GAC<br>54. GAA  **E**-I<br>55. GAG | 60. GGU<br>61. GGC  **G**-II<br>62. GGA<br>63. GGG | U<br>C<br>A<br>G |

**Table 1.** The Table of the standard genetic code (GCT). Total codon space is divided into three parts in correspondence with the harmonic mean (H) of the whole codon space sequence (a) and its half (b), where a = 63, b = 31.5 and H = 42. Bold and underlined amino acids are those that are at the end of a sequence (outer AAs), and they are different from most AAs in the sequence. In the central area, the three stop codons (CT, codon terminations) are crossed out.

## 2. Polarity of amino acids

As there are 63 intervals between 64 points (64 codons), the harmonic mean, calculated by the formula $H(a, b) = 2ab / (a + b)$, (a = 63, b = 31.5), is positioned on the number 42, i.e. on the position of the UGA stop codon. This enables us to formulate *a working hypothesis*, according to which, for better understanding of the relations within the GCT, it is necessary to divide the total codon space into three parts: 0-21, 22-42 and 43-63. This proves that the relation between the number 21 and number 42 (the 42 as harmonic mean, H = 42 in the sequence 0-63) appears to be a realization of "the symmetry in the simplest case" as the ratio 1:2 (Marcus, 1989, p. 103). In Table 1, the first part and the third part of the space are presented in dark tones and the second one in lighter tones.

Indeed, only through such a demarcation of the codon space, can we perceive a characteristic inter-relation between the arrangements of polar and non-polar amino acids (AAs). [Polarity is here taken as hydropathy (Kyte & Doolittle, 1982); more precisely, as the hydropathy index: non-polar amino acids are marked by positive and polar by negative index values (Remark 1).] So, we first perceive that on the right side (in the third part of codon space), all AAs are polar. However, on the left side (the first shaded area) both are present – nonpolar and polar AAs, but in an order imposed by a specific logic. There is, in fact, a spatial sequence (independent of the ordinal numbers of the codons) of non-polar AAs, and only at the end of the sequence (at only one end!), two polar AAs (S and P) have been "added":

[(V–M–I–L–F) – (**S**–**P**)]. (In a linear reading within a linear sequence: all AAs in the first column of Table 1 are non-polar, and first two AAs in second column are polar.) In addition, from a chemical point of view, the added AAs are "the first possible cases": serine as the first possible oxygen derivative, and proline as the first possible cyclic molecule (iso-propyl group bound, by two ending carbon atoms, to the amino acid functional group).

**Remark 1.** The hydropathy index: R = –4.5, K = –3.9, D = –3.5, E = –3.5, N = –3.5, Q = –3.5, H = –3.2, P = –1.6, Y = –1.3, W = –0.9, S = –0.8, T = –0.7, G = –0.4; A = + 1.8, M = + 1.9, C = + 2.5, F = + 2.8, L = + 3.8, V = + 4.2, I = + 4.5 (Kyte & Doolittle, 1982).

The same logic, but with the opposite chemical meanings, applies to the central area (marked by lighter tones). Here, there appears a spatial sequence of polar AAs and in this sequence the three non-polar AAs are also added at the ends (at both ends!); V and A at the beginning, and C at the end of the spatial sequence: [(**V–A**) – (T–Q–H–Y) – (**C**)]. In addition, from a chemical point of view, the added AAs here are also "the first possible cases": valine as the first possible semi-cyclic molecule (cyclo-propyl group bound, by the central carbon atom, to the amino acid functional group), alanine as the first possible carbon derivative, and cysteine as the first possible sulfur derivative. [Carbon, oxygen and sulfur, as derivatives in relation to glycine as the simplest possible amino acid: the substitution of H atom with corresponding functional groups (CH3, OH, SH), respectively. Only nitrogen has remained, as a possible derivative. However, the first possible nitrogen derivative (– CH2–NH2) is not a constituent of the genetic code, but the fourth, in the form of lysine (K) molecule. (For more details of the relationship between valine and proline, in terms of the binding of the isopropyl group, see: Rakočević & Jokić, 1996).]

In connection with polarity of the AAs it is necessary to see their positions in the GCT also from the aspect of "cloister energy" (Swanson, 1984), as well as from the aspect of "polar requirement" (Woese, 1966; Konopel'chenko & Rumer, 1975). Thus, from the aspect of cloister energy, glycine (G) and tryptophan (W) are not polar, but nonpolar AAs. [Glycine (G) is nonpolar also from the aspect of polar requirement.] According to one kind of spatial reading of amino acid sequence (a "diagonal" reading with the last amino acid G on the right in Table 1), in the second shaded area, such a status is expected; expected from the aspect of "sequence logic", presented above. Namely, a possible reading through a spatial sequence is as follows: **W**-(R-N)-(S-K)-R(D-E)-**G**. In such a case, nonpolar AAs are at the ends – at both ends; more exactly, at the beginning and at the end. [At the 43$^{rd}$ codon position there is a nonpolar AA (W); at the 44-59$^{th}$ positions there are polar AAs and at the 60-63$^{rd}$ positions there is a nonpolar AA (G).] In addition, from a chemical point of view the AAs added at the ends are two extreme cases: glycine, as the one and only amino acid without carbon atom in the side chain; tryptophan (W), as the one and only amino acid with two aromatic rings in the side chain.

As for "polar requirement", not only glycine but proline as well is nonpolar and at the same time an extreme amino acid: one and only alicyclic molecule within the set of 20 AAs; as such it is added at the end of the sequence [(V–M–I–L–F) – (**S**–**P**)] within the first shaded area.

## 3. Two classes of AAs handled by two classes of enzymes

The same logic (or similar logic) of "adding at the end" is also valid for the arrangement of AAs, classified into two classes, corresponding to two classes of enzymes aminoacyl-tRNA synthetases (Tables 2.1 & 2.2 in relation to Tables 3.1 & 3.2). It can be said that the corresponding analysis, which was once presented by R. Wetzel (Wetzel, 1989; Rakočević, 1997a), is now shown in a new, clearer light. Bearing this in mind, one must notice that the distribution of AAs into two classes, consistent with enzyme-synthetases classes, corresponds to the hierarchy of their molecule sizes, without any exception (*m* and *v* in Table 2.1): within 10 amino acid pairs, larger molecules (+) belong to class I and the smaller (–) to the class II. On the other hand, two systems – "20" (Table 2.2) and "61" (Table 3.2) amino acid molecules – appear to be in very symmetrical and balanced relationships (Box 1 & 2).

| h | + 0.825 | + 0.943 | + 0.680 | + 0.043 | + 0.251 | + 0.738 | + 0.943 | – **0.000** | + 0.878 | – 0.880 |
|---|---|---|---|---|---|---|---|---|---|---|
| v | + 36 | + 46 | + 30 | + 41 | + 47 | + 52 | + 46 | + 70 | + 83 | + 69 |
| m | + 117.15 | + 131.18 | + 121.16 | + 147.13 | + 146.15 | + 149.21 | + 131.18 | + 174.20 | + 204.23 | + 181.19 |
| I | $V_{10}$ | $L_{13}$ | $C_{05}$ | $E_{10}$ | $Q_{11}$ | $M_{11}$ | $I_{13}$ | $R_{17}$ | $W_{18}$ | $Y_{15}$ |
| II | $G_{01}$ | $A_{04}$ | $S_{05}$ | $D_{07}$ | $N_{08}$ | $T_{08}$ | $P_{08}$ | $K_{15}$ | $H_{11}$ | $F_{14}$ |
| m | 75.07 – | 89.09 – | 105.09 – | 133.10 – | 132.12 – | 119.12 – | 115.13 – | 146.19 – | 155.16 – | 165.19 – |
| v | 03 – | 14 – | 21 – | 30 – | 36 – | 32 – | 31 – | 58 – | 50 – | 62 – |
| h | 0.501 – | 0.616 – | 0.359 – | 0.028 – | 0.236 – | 0.450 – | 0.711 – | 0.283 + | 0.165 – | **1.000** + |

**Table 2.1.** Two classes of amino acids handled by two classes of enzymes. (Class II with 81 and Class I with 123 atoms.) The ten amino acid pairs, *natural* pairs from the chemical aspect, are classified into two classes. Class I contains larger amino acids (larger within the pairs), all handled by class I of enzymes aminoacyl-tRNA synthetases. Class II contains smaller amino acids, all handled by class II of synthetases. Within the rows the given values for the parameters are the following: molecule mass (m), volume of amino acid molecules in angstroms (v) and hydrophobicity (h) on a natural scale (0-1). [Molecule mass and hydrophobicity after: Black & Mould, 1991; volume after: Swanson, 1984.] The order follows the number of atoms within side chains of class II AAs (given here as index); from left to right: first there are aliphatic, and then aromatic AAs. The largest aliphatic molecule of class I (R) is of a zeroth hydrophobicity (h = 0), while the smaller aromatic molecule, F (smaller within the pair F-Y) is of a maximal value (h = 1). [Notice that the pair F-Y is simpler as only aromatic and H-W is more complex as aromatic heterocyclic.]

As we can see, in the first area of the codon space there are AAs from the first class, with three AAs from the second class, added at the end (at one end): [(V–M–I–L) – (**F–S–P**)]; all three as the first possible cases: phenylalanine as the first possible aromatic derivative[2], serine and proline in the manner as we explained above. The opposite situation applies to the right side (second shaded area). As added (at both ends) there are three AAs of the first class, and AAs of the second class are between them; all together: {(**W**, **R**) – [(S, N), (K, R), (G, D)] – **E**)}. (There is also a spatial "diagonal" reading here, but with the last amino acid **E** on the left in Table 1, instead of G as was the case in polarity, in the sequence, shown above.) Thus, all three added AAs are – "the last cases", i.e. the extreme cases: tryptophan as the only amino acid with two rings (within the set of four aromatic AAs), arginine as more complex in the set of two amino derivatives (K, R), and glutamic acid as more complex in the set of two carboxylic AAs (D, E). [This logic is valid for all ten amino acid pairs: more complex (larger) molecule is handled by the first, while the simpler one - by the second class of aminoacyl-tRNA synthetases.]

---
[2] If we know that phenylalanine belongs to a set of 16 amino acids of alanine stereochemical type (with a $CH_2$ group between amino acid "head" and "body"), it becomes clear why a toluene and not benzene derivative must be 'the first possible case'. [About four stereochemical types see: (Popov, 1989; Rakočević and Jokić, 1996).]

| Atom number of 20 AAs (AAs positions in Table 2.1) | | | |
|---|---|---|---|
| Class | Odd | Even | Sum |
| I | 57 | 66 | → 123 |
| II | 33 | 48 | → 81 (9 × 9) |
|  | 90 | 114 | → (10 × 9), (123 – 9) |
|  | Odd | Even | Sum |
| I | 349 | 412 | → 761 (661+100) |
| II | 212 | 282 | → 494 (594-100) |
|  | 561 | 694 | → (661-100), (594+100) |
| (661 – 628 = 33), ( 627 – 594 = 33) | | | |
| 628+627 = 1255 (total nucl. numb. in side chains) | | | |
| Nucleon number of 20 AAs | | | |

**Table 2.2.** Balances of the number of particles within 20 AAs. The balances are given here concerning the relations between parameters in Table 2.1. For atom number: being 9 × 9 in 10 molecules of class II and 10 × 9 at odd positions for both classes; there are differences of 9 atoms etc. For nucleon number: through a specific balance at odd/even positions the difference of 33 atoms appears in relation to middle value (627-628); one balance more: the relations to the number 594 are given and this number also represents the number of atoms within 61 amino acid molecules (cf. Table 3.2).

| a | + 40 | + 78 | – 10 | + 20 | + 22 | – 11 | + 39 | + 102 | – 18 | + 30 |
|---|---|---|---|---|---|---|---|---|---|---|
| I | $4V_{10}$ | $6L_{13}$ | $2C_{05}$ | $2E_{10}$ | $2Q_{11}$ | $1M_{11}$ | $3I_{13}$ | $6R_{17}$ | $1W_{18}$ | $2Y_{15}$ |
| II | $4G_{01}$ | $4A_{04}$ | $6S_{05}$ | $2D_{07}$ | $2N_{08}$ | $4T_{08}$ | $4P_{08}$ | $2K_{15}$ | $2H_{11}$ | $2F_{14}$ |
| a | 4 – | 16 – | 30 + | 14 – | 16 – | 32 + | 32 – | 30 – | 22 + | 28 – |

**Table 3. 1.** Atom number within 61 amino acid molecules. Two classes of AAs as in Table 2.1, except that they are relating only to the number of atoms; (**a**) Number of atoms within codon corresponding 61 amino acid molecules; the number of atoms within amino acid side chains, per "sets" (V = 4, L = 6 etc.), is such that (+/–) designations, i.e. quantities, follow a logic of the strict symmetry: one minus follows two pluses, then two pluses follow one minus etc., for aliphatic AAs; and one plus follows one minus for aromatic AAs.

Through two positions in third part of Table 1, the arginine (R) appears to be the only exception, which "spoils" the presented logic. Arginine also appears as such an exception in the same way as in the Codon Path Cube, as we have previously shown. Namely, in Codon Path Cube two classes of AAs, handled by two classes of enzymes aminoacyl-tRNA synthetases, are splitting into two separated areas, with only one exception, the exception of arginine (Rakočević, 1997a).

In the second part of codon space (in the central codon-space area) the logic is as follows: in the lower part of the space, in the sequence: T, A, **V**, the "different" molecule is the last one (**V**), whereas in the upper part [C–Y–(**H**)–Q] it is – the first to the last (**H**). As we know, this amino acid is also a special case with the value H = 0 in cloister energy. From both these specificities follows that Y – C are at the end of whole amino acid sequence.

| Molecule number | | | |
|---|---|---|---|
| Class | Odd | Even |  |
| I | 12 | 17 | → **29** |
| II | 18 | 14 | → **32** |
|  | **30** | **31** | (29, 30, 31, 32) |
| Class | Odd | Even |  |
| I | 129 | 241 | → **370** |
| II | 104 | 120 | → 224 |
|  | 233 | 361 | → **594** |
| Atom number | | | |

**Table 3. 2.** Balances of the number of particles within 61 amino acid molecules, given here concerning the relations between parameters in Table 3.1. The total atom number within 61 amino acid molecule (594) appears to be in relation to the Shcherbak's "Prime Quantum 37". On the other hand, the number of molecules (through odd/even positions) makes a segment of natural numbers series: 29, 30, 31, 32.

| Box 1 | | | | | |
|---|---|---|---|---|---|
| 20 AAs | → | 9<u>0</u> (Odd) | – 8<u>1</u> (Class II) | = | 9 |
| 61 AAs | → | 233 (Odd) | – 224 (Class II) | = | 9 |
| | | 110 | 110 | | |
| 20 AAs | → | 123 (Class I) | – 114 (Even) | = | 9 |
| 61 AAs | → | 37<u>0</u> (Class I) | – 36<u>1</u> (Even) | = | 9 |

| Box 2 | | | |
|---|---|---|---|
| 20 AAs | | 61 AAs | |
| 90 (Odd) | + | 361 (Even) | = 451 |
| 81 (Class II) | + | 370 (Class I) | = 451 |
| 114 (Even) | + | 233 (Odd) | = 347 |
| 123 (Class I) | + | 224 (Class II) | = 347 |

| | |
|---|---|
| 3A + 3T + D + K + F = 72 | |
| 3V + 2I + C + Q = 72 | |
| L + M + E + R + Y = 66 | |
| **A + T + D + K + F** = 48 (18) | |
| **V + I + C + Q** = 39 | |
| G + P + S + N = 22 (17) | |
| 48 + 39 = 87 | H + W = 29 |
| 66 + 22 = 88 | 29 -18 = 11 |
| | 18 – 17 = 01 |

**Box 1.** This arrangement presents the relationships between 20 and 61 amino acid molecules in the form of atom number difference in the sets of AAs at odd / even positions and / or within the class I / II, of the systems presented in Tables 2.1 and 3.1.

**Box 2.** The same (*mutatis mutandis*) arrangement as in Box 1, but through "plus" instead of "minus" operations. Most AAs on the odd/even positions and/or within the class I/II are the same, but some are diferent (thus, the balance and symmetry are realized) as it is presented on the right side. [The different AAs are: A, T, D, K, F from the first class, and V, I, C, Q from the second one.]

## 4. Arithmetical balances

All presented distinctions, at the same time logical and chemical, are accompanied by the appropriate arithmetical balances (Table 4). Thus, in the first part of the system (the shaded left side) there are 87 atoms in the 8 amino acid molecules, while on the right side there are 98 atoms in 9 amino acid molecules; the difference: for molecules 01 and for atoms 11. In the second (the lighter shaded) part there are 64 atoms: in the middle area (column "A", Table 1) there are 37, while within two ending areas (columns "U-C" and "G" in Table 1) there are 27 atoms; the difference 37 – 27 is 10 (11- 01). [The atom number in column "A": Y15 + H11 + Q11 = 37; in columns "U-C": V10 + A04 + T08 and in column "G": C05; all together equals 27 atoms.]

These results follow if they are calculated only for the amino acid side chains. If, however, the amino acid functional groups (the "heads" of molecules) are taken into account (9 atoms per AA), then instead of 87 and 98 there are 159 and 179 atoms, respectively, which means ± 10 in relation to the arithmetic mean, which is 169. On the other hand, based on the relations to the number of atoms in the second part (127 atoms) and on the differences 159-127 = 2 × 16 and 179-127 = 2 × 26, numbers 16 and 26 appear as the key determinants of the codon distribution (per amino acid) into four classes, based on the four types of amino acid diversity (Rakočević, 2011: Table 4, Eq.4 and Fig. 3 on pp. 826-828; arXiv:1107.1998 [q-bio.OT]).

Besides the balances shown above, there is a balance of the number of atoms in the corresponding codons, and in their nucleotides, but only when the stop codons are excluded (cf. Remark 2). Thus, on the left side there are 2321, whereas on the right side there are 2322 atoms (row "H", Table 4). The difference is exactly one atom (2322 – 2321= 1). In the central area, within the pyrimidine nucleotides (UMP and CMP) there is 971-2, and in the purine ones (AMP and GMP) there is 972 +2 of atoms. A similar balance appears to the number of hydrogen bonds, in the nucleotide bases: there are 161 on the left, and 162 on the right side. In the central area, in the pyrimidine nucleotides, (within their U and C bases) there are 68+5, and 69-5 in the purine type of nucleotides (within their A and G bases).

With the breakdown of the whole codon space into three parts, there inevitably appears one additional amino acid molecule, valine, resulting in the total of 24 amino acids within 16 quadruplets. (Notice that the side chain of valine represents a crossroads between openness and cyclicality.) Only this insight enables us to see new balances, for example, the balance of the number of atoms as well as of isotopes in just 24 molecules, placed into odd-even positions, counted – one at a time or two at a time (Tables 5 & 6).

**Remark 2.** Within U, C, A, G bases there are 12, 13, 15, 16 atoms, respectively. Within nucleotides: UMP, CMP, AMP, GMP → (12, 13, 15, 16), plus 20 atoms in ribose molecule and 8 atoms in phosphoric acid, minus 6 atoms in 2 water molecules; all together equals: UMP-34, CMP-35, AMP-37, GMP-38. This implies that the two inner as well as the two outer columns of GCT (Table 1) contain 3456 of atoms each (including three stop codons) (cf. Remark 3). Thus, within whole GCT there are 3456 + 3456 = 6912. In order to interpret the three parts, determined by the harmonic mean (Table 1), this sum must be reduced by as much as 326 atoms in the three stop codons: 6912-326 = 6586, and this number is actually the product of the Shcherbak's "Prime Quantum 37" (6586 = 37 × 178). [About Shcherbak's "Prime Quantum 37", see in (Shcherbak, 1994), and its geometrical interpretation, see in (Mišić, 2011).] In relation to the row H, in Table 4, we have: 2321+1943+2322 = 6586 (1943 = 969 + 974).

**Remark 3.** The "stop" codons are not the same in standard and mitochondrial Genetic codes. On the other hand, the mitochondrial code, in a way, is more symmetrical than the standard one; and thus more adequate, for example, for p-adic mathematics determination (Dragovich & Dragovich, 2010). Hence, in further research it is necessary to check whether it is possible that the mitochondrial genetic code Table – via the harmonic mean – can also be divided into three equal parts; i.e. are we any closer to answer the question which code is more original, and which occurred in the course of evolution.

|   | I | II | III |
|---|---|---|---|
| A | 8 | 7 | 9 |
| B | 87 | 64 | 98 |
| C | 159 | 127 | 179 |
| D | 22 | 18 | 21 |
| E | 66 | 54 | 63 |
| F | 50/16 | 28/26 | 15/48 |
| G | u 28/9 a<br>c 22/7 g | u 11/14 a<br>c 17/12 g | u 6/21 a<br>c 9/27 g |
| H | 2321 | u,c<br>971-2<br>972+2<br>a,g | 2322 |
| K | 161 | u,c<br>68+5<br>69-5<br>a,g | 162 |

**Table 4.** Distributions and distinctions within three parts of GCT. Roman numerals I, II and III correspond to the three shaded areas in Table 1. A. Number of molecules; B. Number of atoms in the side chains of amino acids; C. Number of atoms in the entire amino acid molecules: amino acid functional groups plus side chains ("head" + "body"); D. Number of codons in the three shaded areas; E. Number of nucleotides; F. Number of pyrimidine/purine nucleotides; G. Number of UMP, CMP, AMP and GMP nucleotides, respectively; H. Number of atoms in nucleotides; K. Number of hydrogen bonds.

| F | 14 | 11 | H | F | 27 | 22 | H |
|---|---|---|---|---|---|---|---|
| L | 13 | 11 | Q | L | | | Q |
| L | 13 | 08 | N | L | 26 | 23 | N |
| I | 13 | 15 | K | I | | | K |
| M | 11 | 07 | D | M | 21 | 17 | D |
| V | 10 | 10 | E | V | | | E |
| V | 10 | 05 | C | V | 15 | 23 | C |
| S | 05 | 18 | W | S | | | W |
| P | 08 | 17 | R | P | 16 | 22 | R |
| T | 08 | 05 | S | T | | | S |
| A | 04 | 17 | R | A | 19 | 18 | R |
| Y | 15 | 01 | G | Y | | | G |
|   | 60 | 65 | **125** |   | 64 | 61 | **125** |
|   | 64 | 60 | **124** |   | 60 | 64 | **124** |
|   | **124** | **125** |   |   | **124** | **125** |   |

**Table 5.** Number of atoms in the side chains of "24" amino acid molecules. The sequence of amino acid molecules is given in Table 1: on the left side one by one at odd/ even positions; on the right side two by two. The result is the same: 124 and 125 atoms in either horizontal or vertical reading. That means that the number of atoms in the first twelve molecules equals the number of atoms in 12 molecules at even positions (124); and, the number of atoms in the last twelve molecules equals the number of atoms in 12 molecules at odd positions (125).

| | | | | | | |
|---|---|---|---|---|---|---|
| F | 28 | 22 | H | F | 54 | 45 | H |
| L | 26 | 23 | Q | L | | | Q |
| L | 26 | 17 | N | L | 52 | 47 | N |
| I | 26 | 30 | K | I | | | K |
| M | 24 | 16 | D | M | 44 | 38 | D |
| V | 20 | 22 | E | V | | | E |
| V | 20 | 12 | C | V | 31 | 48 | C |
| S | 11 | 36 | W | S | | | W |
| P | 16 | 34 | R | P | 33 | 45 | R |
| T | 17 | 11 | S | T | | | S |
| A | 08 | 34 | R | A | 39 | 36 | R |
| Y | 31 | 02 | G | Y | | | G |
| | 122 | 135 | **257** | | 131 | 128 | **259** |
| | 131 | 124 | **255** | | 122 | 131 | **253** |
| | | | **253** | | | | **259** |
| | | | **259** | | | | **253** |

**Table 6.** Number of isotopes in side chains of "24" amino acid molecules. Amino acids are counted in the same way as in the previous Table (Table 5). The result shows the following number sequences 253-255-257-259 and 253-259/253-259 of the number of isotopes (in correspondence with continuity and minimum change principle). The number of isotopes is calculated as follows (in the example for serine): [(3 × 2) H] + [(1 × 2) C] + [(1 × 3) O] = 11. (These are the only stable isotopes, i.e. stable nuclides: 2 for hydrogen, 2 for carbon and 3 for oxygen.)

## 5. Hydrophobicity of amino acids

The next parameter which corresponds to the splitting of the GCT (Table 1) into three parts is hydrophobicity ($h$ in Table 2.1). For eight out of ten amino acid pairs everything is the same as in the distribution of AAs in relation to two classes of synthetases. The remaining two pairs, F-Y and K-R, are in a vice versa position (through +/− designations in Table 2.1). The epilogue is that in the shaded area on the left side, there is, again, the same situation as with polarity: [(V–M–I–L–F) – (**S**–**P**)]. Each of the first five amino acids appears to have a higher value for the hydrophobicity within their own pairs. The other two, which are added at the end, have lower values. [In connection with the hydrophobicity, it is important to recall that just this variant of parameter, as a "natural scale" (Black and Mould, 1991), is the best one among a host of other parameters which also measure the hydrophobicity of AAs (Chechetkin and Lobzin, 2011).]

In the shaded area on the right side of Table 1, in this parameter, arginine (R) is not an exception any more. All AAs [R, (S, N), (**K**, R), (G, D)], which are located between the two ends (**W-E**), have lower values within their own pairs; all but one: lysine (K) appears now to be an exception, instead of arginine (R). The rest of two AAs (at two ends) have higher values (W-E).

The same logic of "adding at the end" is valid also for the arrangement of AAs in the middle area if we understand that this area consists of two parts: more exactly two sub-parts: the sub-part up and sub-part down. In the first case we have: (**Q**–H–Y–**C**), where two central AAs have lower, whereas the two AAs at two ends have higher values in their own pairs; and, in the second case (**V**–A–T), the T and A have lower values but V, at the end, possesses a higher one.

## 6. Cube equation as a model for GC

The splitting of the total codon space into three parts: 0-21, 22-42 and 43-63, i.e. the triple multiplication of the number 21, inevitably gives us the idea (the hypothesis for further researches) that, perhaps, it makes sense to do the fourth multiplication, i.e. to add the amino acid space – the 20 amino acid molecules plus one stop signal (stop signal, encoded by three stop codons) to the codon space. In support of this idea, there is the fact that there is such a justification for splitting the set of the constituents of genetic code into three parts that corresponds to general form (model) of a cubic equation (21 × 4 = 84) (Table 7).

Table 7 gives the most general form of a cubic equation. If one of the three possible real solutions for *x* (in the case when a = b = c = 1) takes values from the series of natural numbers (*x* = 1, 2, 3 ...), it is immediately obvious that the fourth case (*x* = 4) is the model (corresponding to the structure) of the genetic code; there are 64 codons, 16 aliphatic and 4 aromatic amino acids, with a total of 84 quantities.

In addition, there are some other regularities. Thus, through validity of the similarity principle and self-similarity principle there is a chemical correspondence between four aromatic amino acids and four nucleotide bases, also of aromatic nature. Moreover, there is a balance in the number of atoms: U12+G16 = C13+A15 = 28; and in the amino acid side chains: F14+Y15 = H(14-3) + W(15+3) = 29. On the other hand, the total of 84 quantities is a "missing link", because, in previous papers, we showed that the GC is determined by a quadratic equation in the form of golden mean (Rakočević, 1998, Table 2, p. 288), as well as by two systems of linear algebraic equations with solutions $x_1$, $y_1$ and $x_2$, $y_2$ (Rakočević, 2011, Eq.4, p. 827). However, this is also the "missing link" from a different point of view. Namely, in a previous work, we showed that the GC is determined by geometric progressions with quotient 2 and 3, respectively (Rakočević, 2011, Tables A.1 & A.2, p. 839), and now it appears that the determinant of GC is still the geometric progression with the quotient of number 4. Finally, the 64 codons exist as eight 8-membered octets (the principle of self-similarity!): YUN-YCN-RUN-RCN / YAN-YGN-RAN-RGN.

**Table 7.** The cube equation as a model for the structure of genetic code. Second case corresponds with eight codon octets (YUN, RUN, YCN, RCN, YAN, RAN, YGN, RGN), four quartets of aliphatic AAs (G-P,V-I), (A-L, K-R), (S-T, C-M), (D-N, E-Q) and two doublets of aromatic AAs (F-Y, H-R). Fourth case corresponds with 64 codons, 16 aliphatic AAs and 4 aromatic AAs. Two last columns: the correspondence with the octal and binary numeral systems, respectively (cf. Tab. B.1).

| $ax^3 + bx^2 + cx + d = 0$ If a = b = c = 1, then: $x^3 + x^2 + x = d$ | Octal | Binary |
|---|---|---|
| If (x = 1, 2, 3, 4, 5, ... ), then: | | |
| 1 + 1 + 1 = 03 | | |
| 8 + 4 + 2 = 14 | $8^1$ | $2^3$ |
| 27 + 9 + 3 = 39 | | |
| **64 + 16 + 4 = 84** | $8^2$ | $2^6$ |
| 125 + 25 + 5 = 155 | | |
| ... | | |

## 7. Outer and inner AAs in GCT

Amino acids that have the status "added at the end" can be designated as *outer*[3] and all other as *inner* (Table 8.1, in relation to Box 3). But also without a division of codon space into three parts, it is self-evident from GCT that eight AAs (F-S-Y-C and V-A-E-G) have the status of "outer AAs". And, as we said in chapter 2, all these AAs are „extreme" from a chemical point of view: glycine (G) as the simplest possible amino acid, alanine (A) as the first possible carbon derivative, serine (S) as the first possible oxygen derivative, cysteine (C) as the first possible sulfur derivative, valine (V) as the first possible semi-cyclic molecule, phenylalanine (F) as the first possible aromatic amino acid, and tyrosine (Y) as its first possible oxygen derivative. The only exception is glutamic amino acid (E), which is not "the first", but "the last" case in the set of only two AAs, which possess carboxylic functional group in the side chain.

---

[3] This logic is valid only for the "ends" in the three shaded parts of Table 1, and not for T and Q which have the status "be at the end" within two sub-parts of the middle lower shaded part of Table 1.

However, only with the division of GCT into three parts, we can see that two AAs more also have the same status ("outer AAs"): triptophan (W) and proline (P); triptophan as the only amino acid with two aromatic rings and proline as the only amino acid, which possesses one non-aromatic ring (Tab. 8.1).

If we compare two sets of AAs in Table 8.1 with two sets in Box 4 and with two sets, i.e. two classes in Table 2.1, it becomes evident that both sets in Table 8.1 contain two or three amino acid doublets and six or four singlets each. Three doublets with **6** AAs and **4** singlet AAs (the third case in Box 3) if AAs are chosen from the system in Table 2.1; two doublets with **4** AAs and **6** singlet AAs (the fourth case in Box 3)[4] if AAs are chosen from the system in Box 4. In such an arrangement the balance of the number of particles is self-evident: either for the number of atoms (Table 8.1)[5] or for the number of protons and nucleons (Box 5).

| | | | | | | |
|---|---|---|---|---|---|---|
| **40** | G | 01 | 10 | V | | |
| | F | 14 | 15 | Y | | |
| (11-1) | A | 04 | 13 | L | | |
| | S | 05 | 08 | *T* | | |
| **50** | *C* | 05 | 11 | M | | |
| | P | 08 | 13 | I | 63 | (102+01) |
| | E | 10 | 07 | D | | |
| | W | 18 | 11 | H | (11+1) | |
| | N | 08 | 11 | Q | 51 | (102-01) |
| | K | 15 | 17 | R | | |
| Odd / Even | | 28 | 64 | | (102-10) | |
| Even / Odd | | 60 | 52 | | (102+10) | |
| | | **(81+7)** | **123-7)** | | | |

**Table 8.1.** The atom number balance between outer and inner AAs. Up and on the left there are outer AAs. Down and on the right there are inner AAs. The choice from the system in Table 2.1, valid for outer AAs: (G-V, F-Y, S-C) plus (A, P, E, W); and for inner AAs: (K-R, N-Q, T-M) plus (L, I, D, H). The choice from the system in Box 4, valid for outer AAs: (G-V, F-Y) plus (S, C, A, P, E, W); and for inner AAs: (K-R, N-Q) plus (T, M, L, I, D, H).

With this knowledge it makes sense to ask the following question: which amino acid pairs are present in only one of the three areas and which are present in two areas (first pair-member in one, and second pair-member in another area). With the answer to this question we have a new insight in a perfect balance of the number of atoms within the set of 16 aliphatic AAs as well as within the set of all 20 AAs (Table 8.2). Therefore, three pairs are present in only one area (D-E, P-I, K-R), and five pairs are present in two different areas (G-V, A-L, S-C, T-M, N-Q). As for two aromatic pairs (such pairs as in Table 2.1: F-Y and H-W), they are present in two different areas each. However, as we can see from Table 8.2, for a full balance, the amino acid pairing must be different, but also with a chemical justification: F-W and H-Y (F-Y because Y is an oxygen derivative of F; F-W because both contain the original benzene ring; H-W because both AAs are heterocyclic and H-Y because both AAs are more polar than F-W). With this change there is also a balance in the number of molecules: 5-1 molecule pairs are present in GCT within only one area (of three possible areas) (D-E, P-I, K-R, **H-Y**), and 5+1 molecule pairs are present in two different areas (G-V, A-L, S-C, T-M, N-Q, **F-W**).

---

[4] The splitting of 10 amino acid pairs into 5 ± 1 doublets and singlets, in this way, appears to be the best solution and it is in accordance with "the symmetry in the simplest case" (Marcus, 1989).

[5] With insight into the results, shown in Tables 8.1 and 8.2, one is forced to propose a hypothesis (for further researches) that here, there really is a kind of *intelligent design*; not the original intelligent design, dealing with the question - intelligent design or evolution (Pullen, 2005), which is rightly criticized by F.S. Collins (2006). Here, there could be such an intelligent design, which we could call "Spontaneous Intelligent Design" (SPID) that is consistent with that design which was presented by F. Castro-Chavez (2010), and is also in accordance with the Darwinism. [F. Castro-Chavez (2010, p. 718): "We can conclude that the genetic code is an intelligent design that maximizes variation while minimizing harmful mutations."] Actually, it can be expected that the hypothetical SPID, contained in the results shown in Table 8.1, is in accordance with an identical (or similar?) SPID, presented in the only diagram, in Darwin's book "Origin of Species" (Darwin, 1996), as we have shown through an analysis of that diagram in one of our books (Rakočević, 1994; www.rakocevcode.rs). [In the case of the statement that *spontaneity* and *intelligent design* are mutually opposite, one must ask the question: isn't it true that human intelligence is the result of a spontaneous evolutionary process?]

| G | 01 | 10 | V | G | 01 | 10 | V |
|---|---|---|---|---|---|---|---|
| A | 04 | 13 | L | A | 04 | 13 | L |
| S | 05 | 05 | C | S | 05 | 05 | C |
| T | 08 | 11 | M | T | 08 | 11 | M |
| N | 08 | 11 | Q | N | 08 | 11 | Q |
|   |   |   |   | D | 07 | 10 | E |
| D | 07 | 10 | E | P | 08 | 13 | I |
| P | 08 | 13 | I | K | 15 | 17 | R |
| K | 15 | 17 | R | H | 11 | 15 | Y |
|   |   |   |   | F | 14 | 18 | W |
|   | 22 | 51 | **73** |   | 33 | 69 | **102** |
|   | 34 | 39 | **73** |   | 48 | 54 | **102** |
|   |   |   |   |   | 81 | 123 | **204** |

**Table 8.2.** The distribution of amino acid pairs into three parts of GCT. On the left: first five aliphatic amino acid pairs appear to be together each, within one of the three areas in Table 1. The rest of three pairs are impaired: first member in one area and the second one in another area. On the right: the 8 aliphatic plus two aromatic pairs of AAs. The atom number balance is evident in both halves of the system.

In Table 8.1, up and left, there are the AAs with the status "added at the end" (outer) within the three parts of GCT (Table 1), while all the others are down and on the right (inner). Two columns correspond to two rows in Table 2.1. In the first column there are AAs handled by class II of aminoacyl-tRNA synthetases, all but three – C, EW – from the first class; in the second column there are those AAs which are handled by class I aminoacyl-tRNA synthetases, except three – T, DH – from class II.

The balances of the number of atoms in Table 8.1 are directly indicated. However, there is a hidden balance between the number of atoms within two columns. Originally, first column (first row – Class II – in Table 2.1) contains 81 atoms, which means 102 – 21, where 102 is the arithmetic mean, while the second column contains 123, i.e. 102 + 21 atoms. Furthermore, the difference 123 – 81 = 42, as a quantity "42", corresponds to quantity "42" which also appears to be harmonic mean of the codon space 0-63 (see Chapter 2).[6]

**Remark 4.** The proton number within amino acid side chains: (**G = 01, V = 25; F =49, Y = 57**), (**A = 09, S = 17, C = 25, P = 23, E = 39, W = 69**) / (L = 33, T = 25, M = 41; I = 33, D = 31, H = 43), (N = 31, Q = 39; K = 41, R = 55).

**Remark 5.** The nucleon number within amino acid side chains: (**G = 01, V = 43; F =91, Y = 107**), (**A = 15, S = 31, C = 47, P = 41, E = 73, W = 130**) / (L = 57, T = 45, M = 75; I = 57, D = 59, H = 81), (N = 58, Q = 72; K = 72, R = 100).

| Box 3 | | Box 4 | |
|---|---|---|---|
| Singlets | Doublets | a | b |
| inner + outer | inner + outer | | |
| 10 + 10 (5 + 5) | 0 + 0 (5 - 5) | A – L | G – V |
|  |  | S – T | P – I |
| 8 + 8  (5 + 3) | 1 + 1 (5 - 3) | C – M |  |
|  |  | D – E |  |
| **6 + 6  (5 + 1)** | **2 + 2 (5 - 1)** | N – Q |  |
|  |  | K – R |  |
| **4 + 4  (5 - 1)** | **3 + 3 (5 + 1)** | H – W |  |
|  |  | F – Y |  |
| 2 + 2  (5 - 3) | 4 + 4 (5 + 3) |  |  |
| 0 + 0  (5 - 5) | 5 + 5 (5 + 5) |  |  |

**Box 3.** Possible "choices" of amino acid singlets or doublets from the pairs given in Table 2.1 are presented here. In Table 1 they take outer or inner positions; that means that in some cases AAs appear separately, in other cases they appear in pairs, as shown in Table 8.1. If we compare Box 3 with Table 8.1, it is clear that the two central cases are "chosen".

**Box 4.** Going from the system presented either in Table 2.1 or in Box 4, to the system in Table 8.1, the amino acid pairs can be split

---
[6] The splitting of atom number space into three parts (3 × 14 = 42) appears to be a simulation of such a splitting of codon space also into three parts (3 × 21 = 63).

into AAs as singlets or doublets, in the framework of these possibilities. Column *a* as in Survey 1.1 and column *b* as in Survey 1.2, both in: Rakočević and Jokić, 1996. In column *a* there are AAs from the alanine stereochemical type, while in column *b* there are AAs from three non-alanine stereochemical types: glycine type (G), proline type (P) and valine type (V-I).

---

**Box 5.** Number of protons and nucleons within outer and inner AAs within the system in Tab. 8.1

**Proton number**

Number of protons in outer AAs: Singlets (18**2**) – Doublets (13**2**) = 50
Number of protons in inner AAs: Singlets (20**6**) – Doublets (16**6**) = 40
Difference of differences: 50 – 40 = 10

Positions in both columns:  36**8** – 31**8** = 50
Even + Odd  = 368
Odd + Even  = 318

**Nucleon number** (Differences: ± 20)
Outer / Inner AAs:        57**9** / 67**6**
Left / Right column:      55**9**/ 69**6**

---

**Box 5.** All data are related to the system in Table 8.1

**Table. 9.** The harmonic mean of harmonic mean. Harmonic mean of 63 and its half is H = 42. The half of 42 is H/2 = 21. Finely, harmonic mean of 21 and its half is h = 14 (cf. Table B.1).

| n | $2^n$ | Intervals | H/2 | h/2 |
|---|---|---|---|---|
| 1 | 2 | 0 – 1 | | |
| 2 | 4 | 0 – 3 | 1 | |
| 3 | 8 | 0 – 7 | | |
| 4 | 16 | 0 – 15 | 5 | |
| 5 | 32 | 0 – 31 | | |
| 6 | 64 | 0 – 63 | 21 | (7) |
| 7 | 128 | 0 – 127 | | |
| 8 | 256 | 0 – 255 | 85 | |
| 9 | 512 | 0 – 511 | | |
| 10 | 1024 | 0 – 1023 | 341 | |
| 11 | 2048 | 0 – 2047 | | |
| 12 | 4096 | 0 – 4095 | 1365 | (455) |
| ... | | | | |

The changes caused by a direct EW/DH "crossing-over", and an indirect (diagonal) C/T "crossing-over", the atom number within columns becomes $102 \pm 14$, respectively, where 14 is the harmonic mean of the number 21 and its half (h = 14, in relation to Table 9). Thus, bearing in mind that (102 +14) - (102 -14) = 28, where 28 is the harmonic mean of 42, it is self-evident that the relationships in Table 8.1 (results 81 + 7 and 123 − 7) correspond with the 6th case in the number arrangement in Table 9. In other words, the EW/DH and C/T "crossing-overs" occur correspondingly with the harmonic mean of the harmonic mean (H = 42 and 2h = 28; "H" and "h" in relation to the number arrangement in Table 9). [Cf. shaded rows with  columns in Table B.1.]

## 8. Three parts of GCT by the codons

In addition to previously presented large amount of regularities, which justify the codon space division into three equal parts, it makes sense to examine the validity of this division from the aspect of relations between the codons, as well. Beside other possible relations, here we present only those relations that are in connection with the nucleotide number interdependence − relations of lower and higher molecule complexity − in the three specified parts ("0" for lower and "1" for higher molecule complexity in binary records, given inTables A.1, A.2 and A.3 in Appendix A). In addition, two pyrimidines in relation to two purines are of lower complexity, as well as two nucleotides with two hydrogen bonds in relation to two nucleotides with three hydrogen bonds: (U, C → A, G), (U, A → C, G) .

The hierarchy which we have just presented follows the original concept of R. Swanson (1984), which means that the hierarchy is derived from the number of units in binary codon record, in the Gray code model, on the binary tree as well as in the GCT (Table 1).

As we can see from Tables 10 & 11 (in relation to the Appendix A), on the 6-bit binary tree, i.e. within 64 words in GCT there are exactly 6 +1 patterns that are related to the occurrence of a unit (a nucleotide of higher complexity) in binary records (column *a* in Table 10 in relation to Table 11). In the six patterns it is inevitable for such a change to occur i.e. it is inevitable that at least one nucleotide of higher complexity appears.[7] However, there is exactly one pattern (the seventh) in which there is no hierarchy change, a nucleotide of higher complexity does not appear, and that is zeroth codon UUU (000000) (cf. "0" in column "$a_1$" of Table 10 and "F" in column "1" of Table 11).

| I | | II | | III | |
|---|---|---|---|---|---|
| $a_1$ | $b_1$ | $a_2$ | $b_2$ | $a_3$ | $b_3$ |
| 0 | (1) | 1 | (1) | 2 | (1) |
| 1 | (5) | 2 | (5) | 3 | (5) |
| 2 | (9) | 3 | (9) | 4 | (9) |
| 3 | (6) | 4 | (5) | 5 | (5) |
| 4 | (1) | 5 | (1) | 6 | (1) |
| | 22 | | 21 | | 21 |

**Table 10.** Codon hierarchy within three parts of GCT. The significance of columns is as follows: a. Number of occurrences of nucleotides of higher complexity (purinees and/or nucleotides with three hydrogen bonds); b. Number of codons in the patterns provided in "a".

| b | 1 | 5 | | | | 9 | | | | | | | 6 / 5 | | | | | 1 |
|---|---|---|---|---|---|---|---|---|---|---|---|---|---|---|---|---|---|---|
| $b_1$ | F | F | L | L | S | I | L | L | L | S | S | P | I | I | V | L | S | P | P | M | V | P |
| $b_2$ | Y | T | Y | * | H | C | V | T | T | A | * | H | Q | C | * | V | T | A | A | Q | - | A |
| $b_3$ | N | R | N | K | D | S | W | R | R | K | D | E | S | R | G | R | E | R | G | G | - | G |

**Table 11**. AAs within three parts of GCT after Table 10. Explanation in the text.

## 9. Concluding remarks

The presented results confirm *the working hypothesis* according to which it is necessary to divide the total codon space within GCT into three parts: 0-21, 22-42 and 43-63, and also the results go in favor of our earlier hypothesis that the genetic code was already complete in prebiotic conditions (Rakočević, 2004). However, to what we have said in previous work about the prebiotic completeness of GC[8], we now add one idea more (hypothesis for further research): only such an aggregation of AAs that can generate all regularities and interrelationships that are presented here can also generate life as such. But, bearing in mind the relationships, presented in Tables B.1, C.1 and C.2, it makes sense to formulate a hypothesis about possible extraterrestrial life. By doing so, we start from the following facts: four nucleotide bases (U, C, A, G) and 18 non-sulfur amino acids are made out of the first possible, i.e. of the simplest non-metals (H, C, N, O).[9] Functional groups in their simplest compounds are also found in the composition of amino acid functional group: [(H–CH$_3$), (H–OH), (H–NH$_2$), (C=O)] → [(H–C(H)–(C=O)–(OH)–(NH$_2$)]. On the other hand, in Tables B.1, C.1 and C.2 we see that there is a strict relation between the number of atoms in 20 amino acids (384) and in 64 codons, i.e. in 192 nucleotides, through an integer multiplication (384 x 018 = 3456 + 3456) (cf. Table C.1 ). Thus, in my opinion, the next hypothesis can be formulated based on all these facts: all the planets in the universe, which possess water (and other conditions needed for life) must necessarily have just the same – the simplest possible – terrestrial genetic code.[10]

---

[7] The sequence of digits in the record excludes the dilemma when, for example "C", should be read as a pyrimidine nucleotide (0), and when, as a nucleotide of higher complexity with three hydrogen bonds (1), which was otherwise specified in the regularities presented by R. Swanson (1984).

[8] "At a later stage many nucleotide/amino-acid aggregations … had been realized … Each of those aggregations could (and must) have its own ''evolution'', but only one could have been selected — the one that gained the characteristic of self-reproduction …" (Rakočević, 2004, p. 232).

Without further justification for making the hypothesis, according to which intelligent beings from outer space have been incorporated "intelligent signals" in the terrestrial genetic code (*sh*Cherbak and Makukov, 2013)[11], we believe that the idea about an artificial genetic code can be useful when it comes to stocking the digital information. In such a case, we believe that the presented regularities, contained in the standard genetic code, would benefit, particularly the building of genetic-code- algorithmic structures.

Besides everything stated above, on the basis of these results, it makes sense to give two predictions:

1. The presented relations between amino acids must be, *mutatis mutandis*, expressed in the structure of proteins, especially in terms of determination of invariant, conservative and radical amino acid positions, with distinct biological effects;

2. The presented unity of form and essence (mathematical regularities versus physicochemical properties), together with such a unity presented in our previous paper (Rakočević, 2011, Eq. 3 on p. 826 in relation to Fig. 3 on p. 828), must be reflected – through genes expression – on the relationship between genotype and phenotype in the development and evolution of organisms; everything in accordance with Futuyma's idea that the lack of knowledge of how genotypes generate phenotypes is the greatest gap in our understanding of evolution processes (Futuyma, 1979).

---

[9] The inclusion of two sulfur AAs occurs through the inclusion of sulfur, the oxygen's first neighbour in the sixth group; a completion of nucleotide molecule through the inclusion of phosphorus, nitrogen's first neighbour in the fifth group of the Mendeleev's Periodic System.

[10] "The simplest possible code" in relation to the following facts: glycine as the first possible amino acid (AA) without carbon in side chain; alanine as the first possible carbon AA with the open side chain; valine as the first possible half-cyclic AA; proline as the first possible cyclic AA; leucine and isoleucine as the first possible branched AAs (first possible isomers); phenylalanine as the first possible aromatic AA; all other AAs as derivatives, realized through the validity of the minimum change principle, continuality principle and self-similarity principle. For example: serine and cysteine in relation to alanine; aspartic AA in relation to serine and glutamic AA in relation to aspartic one (from both these dicarboxilic AAs follow their two amides); lysine with four carbon atoms in side chain in relation to isoleucine, also with four carbon atoms in side chain; tyrosine and tryptophan in relation to phenylalanine etc. From this logic minimally deviate methionine, arginine and histidine, but they, just as such ("clumsy") are necessary for completion the Gaussian algorithm (Rakočević, 2011, Fig. 9).

[11] "As the actual scenario for the origin of terrestrial life is far from being settled, the proposal that it might have been seeded intentionally cannot be ruled out." ( *sh*Cherbak and Makukov, 2013).

## Appendix A

Tables A.1, A.2 and A.3, provide a detailed disposition (arrangement) of amino acids and codons according to Tables 10 & 11. Codons are arranged in the same way (by the same ordinal numbers) as in Table 1, corresponding to the number of units in the binary record of that arrangement within quantum 1-5-9-6-1 in Table A.1 and within quantum 1-5-9-5-1 in Tables A.2 & A.3 (cf. Survey 1, below). In all three cases that and such disposition is accompanied by a strict atom number balance within the amino acid side chains. Partial balance, through quantum combination in sequence 1-5-9-5-1 is self-evident. However, if we observe the three sums within the three parts (Survey 2 in relation to Survey 3), we realize that the possible meaning of the relation is not visible at first sight. The fact that differences "14" and "62" correspond to the second perfect number (28) and to the third perfect number (496), could be, in further researches, understood only as a curiosity, or as an additional proof that perfect and friendly numbers are indeed the determinants of the genetic code (Rakočević, 2007b).

| 1 | 5 | 9 | 5 | 1 | Survey (1) |
|---|---|---|---|---|---|
|  | **4** | **4** | **4** | **4** |  |

| II | III | I |  | Survey (2) |
|---|---|---|---|---|
| 152 | 214 | 228 | $62 \times \mathbf{8} = 496$ |  |
|  | 62 | 14 | $14 \times \mathbf{2} = 28$ |  |

| $4^1$ $2^{\underline{2}}$ | $8^1$ $2^{\underline{3}}$ | $16^1$ $2^{\underline{4}}$ | $32^1$ $2^{\underline{5}}$ | $64^1$ $2^{\underline{6}}$ | Survey (3) |
|---|---|---|---|---|---|
| $4^2$ $2^{\underline{4}}$ | $8^2$ $2^{\underline{6}}$ | $16^2$ $2^{\underline{8}}$ | $32^2$ $2^{\underline{10}}$ | $64^2$ $2^{\underline{12}}$ |  |

| 1 | 5 | | | | | 9 | | | | | | | | | 6 | | | | | | 1 |
|---|---|---|---|---|---|---|---|---|---|---|---|---|---|---|---|---|---|---|---|---|---|
| F | F | L | L | S | I | L | L | S | S | P | I | I | V | | L | S | P | P | M | V | P |
| 0 | 1 | 2 | 4 | 8 | 16 | 3 | 5 | 6 | 9 | 10 | 12 | 17 | 18 | 20 | 7 | 11 | 13 | 14 | 19 | 21 | 15 |
| U | U | U | C | U | A | U | C | C | U | U | C | A | A | G | C | U | C | C | A | G | C |
| U | U | U | U | C | U | U | U | C | C | C | U | U | U | | U | C | C | C | U | U | C |
| U | C | A | U | U | U | G | C | A | C | A | U | C | A | U | G | G | C | A | G | C | G |
| 0 | 0 | 0 | 0 | 0 | 0 | 0 | 0 | 0 | 0 | 0 | 0 | 0 | 0 | 0 | 0 | 0 | 0 | 0 | 0 | 0 | 0 |
| 0 | 0 | 0 | 0 | 0 | **1** | 0 | 0 | 0 | 0 | 0 | 0 | **1** | **1** | **1** | 0 | 0 | 0 | 0 | **1** | **1** | 0 |
| 0 | 0 | 0 | 0 | **1** | 0 | 0 | 0 | 0 | **1** | **1** | **1** | 0 | 0 | 0 | 0 | **1** | **1** | **1** | 0 | 0 | **1** |
| 0 | 0 | 0 | **1** | 0 | 0 | 0 | **1** | **1** | 0 | 0 | **1** | 0 | 0 | **1** | **1** | 0 | **1** | **1** | 0 | **1** | **1** |
| 0 | 0 | **1** | 0 | 0 | 0 | **1** | 0 | **1** | 0 | **1** | 0 | 0 | **1** | 0 | **1** | **1** | 0 | **1** | **1** | 0 | **1** |
| 0 | **1** | 0 | 0 | 0 | 0 | **1** | **1** | 0 | **1** | 0 | 0 | **1** | 0 | 0 | **1** | **1** | **1** | 0 | **1** | **1** | **1** |
| 14 | | | | | | | | | 93 **(114+1)** | | | | | | | | | | | | 8 |
| | | | 58 | | | | | | **(114-1)** | | | | | | | | | 55 | | | |

**Tab. A.1.** Relationships of AAs and codons within first part of GCT after Table 11. Amino acids and codons; the codons with their ordinal numbers and binary records. From left to the right there is a hierarchy of the units (number "1") within binary records of codons. Below, the number of atoms within amino acid side chains and their sums appear to be as balance solutions; within three columns there are 14 + 93 + 8 = 114+1; and within two columns there are 58 + 55 = 114 -1.

| 1 | | 5 | | | | | | 9 | | | | | | | | | 5 | | | | | 1 |
|---|---|---|---|---|---|---|---|---|---|---|---|---|---|---|---|---|---|---|---|---|---|---|
| Y | | T | Y | * | H | C | | V | T | T | A | * | H | Q | C | * | V | T | A | A | Q | A |
| 32 | | 24 | 33 | 34 | 36 | 40 | | 22 | 25 | 26 | 28 | 35 | 37 | 38 | 41 | 42 | 23 | 27 | 29 | 30 | 39 | 31 |
| U | | A | U | U | C | U | | G | A | A | G | U | C | C | U | U | G | A | G | G | C | G |
| A | | C | A | A | A | G | | U | C | C | C | A | A | A | G | G | U | C | C | C | A | C |
| U | | U | C | A | U | U | | A | C | A | U | G | C | A | C | A | G | G | C | A | G | G |
| | | | | | | | | | | | | | | | | | | | | | | |
| **1** | | 0 | **1** | **1** | **1** | **1** | | 0 | 0 | 0 | 0 | **1** | **1** | **1** | **1** | **1** | 0 | 0 | 0 | 0 | **1** | 0 |
| 0 | | **1** | 0 | 0 | 0 | 0 | | **1** | **1** | **1** | **1** | 0 | 0 | 0 | 0 | 0 | **1** | **1** | **1** | **1** | 0 | **1** |
| 0 | | **1** | 0 | 0 | 0 | **1** | | 0 | **1** | **1** | **1** | 0 | 0 | 0 | **1** | **1** | 0 | **1** | **1** | **1** | 0 | **1** |
| 0 | | 0 | 0 | 0 | **1** | 0 | | **1** | 0 | 0 | **1** | 0 | **1** | **1** | 0 | 0 | **1** | 0 | **1** | **1** | **1** | **1** |
| 0 | | 0 | 0 | **1** | 0 | 0 | | **1** | 0 | **1** | 0 | **1** | 0 | **1** | 0 | **1** | **1** | **1** | 0 | **1** | **1** | **1** |
| 0 | | 0 | **1** | 0 | 0 | 0 | | 0 | **1** | 0 | 0 | **1** | **1** | 0 | **1** | 0 | **1** | **1** | **1** | 0 | **1** | **1** |
| 15 | | | | | | | | | | | 57 **(76)** | | | | | | | | | | | 4 |
| | | | | 39 | | | | | | | **(76)** | | | | | | | | 37 | | | |

**Tab. A.2.** Relationships of AAs and codons within second part of GCT after Table 11. The number of atoms within amino acid side chains and their sums appear to be balance solutions; within three columns there are 15 + 57 + 4 = 76; and within two columns there are 39 + 37 = 76. Everything else as in Table A.1.

| 1 | | 5 | | | | | | 9 | | | | | | | | | 5 | | | | | 1 |
|---|---|---|---|---|---|---|---|---|---|---|---|---|---|---|---|---|---|---|---|---|---|---|
| N | | R | N | K | D | S | | W | R | R | K | D | E | S | R | G | R | E | R | G | G | G |
| 48 | | 44 | 49 | 50 | 52 | 56 | | 43 | 45 | 46 | 51 | 53 | 54 | 57 | 58 | 60 | 47 | 55 | 59 | 61 | 62 | 63 |
| A | | C | A | A | G | A | | U | C | C | A | G | G | A | A | G | C | G | A | G | G | G |
| A | | G | A | A | A | G | | G | G | G | A | A | A | G | G | G | G | A | G | G | G | G |
| U | | U | C | A | U | U | | G | C | A | G | C | A | C | A | U | G | G | G | C | A | G |
| | | | | | | | | | | | | | | | | | | | | | | |
| **1** | | **1** | **1** | **1** | **1** | **1** | | **1** | **1** | **1** | **1** | **1** | **1** | **1** | **1** | **1** | **1** | **1** | **1** | **1** | **1** | **1** |
| **1** | | 0 | **1** | **1** | **1** | **1** | | 0 | 0 | 0 | **1** | **1** | **1** | **1** | **1** | **1** | 0 | **1** | **1** | **1** | **1** | **1** |
| 0 | | **1** | 0 | 0 | 0 | **1** | | **1** | **1** | **1** | 0 | 0 | 0 | **1** | **1** | **1** | **1** | 0 | **1** | **1** | **1** | **1** |
| 0 | | **1** | 0 | 0 | **1** | 0 | | 0 | **1** | **1** | 0 | **1** | **1** | 0 | 0 | **1** | **1** | **1** | 0 | **1** | **1** | **1** |
| 0 | | 0 | 0 | **1** | 0 | 0 | | **1** | 0 | **1** | **1** | 0 | **1** | 0 | **1** | 0 | **1** | **1** | **1** | 0 | **1** | **1** |
| 0 | | 0 | **1** | 0 | 0 | 0 | | **1** | **1** | 0 | **1** | **1** | 0 | **1** | 0 | 0 | **1** | **1** | **1** | **1** | 0 | **1** |
| | | | | | | | | | | | (107) | | | | | | | | | | | |
| 8 | | | | 52 | | | | | | | **(107)** | | | | | | | | 46 | | | 1 |

**Tab. A.3.** Relationships of AAs and codons within third part of GCT after Table 11. The number of atoms within amino acid side chains and their sums appear to be as balance solutions; within central column there are 107 atoms as well as within other four columns. All other as in Tab. A.1 and A.2.

**Appendix B**

Table B.1 coresponds with the 4$^{th}$ case in Table 7, the 6$^{th}$ case in Table 9 and with the second case in Survey 3 (App. A). The multiplies of binary quantums (2 exp *n*, where *n* takes the values from series of even natural numbers) show a specific connection between numbers 6, 64 and 384, and these numbers are significant for the genetic code. [Cf. quantum 384 in Table B.1 with the Plato's harmonic number 384 (in Timaeus), as it is presented in our previous paper (Rakočević, 2011, Appendix, Table A.2) (Cf. also Table B.1 with Table 9.]$^{12}$

| | | | | | | |
|---|---|---|---|---|---|---|
| $2^2$ | x 2 | = | 4 | x 2 | = | 8 |
| $2^4$ | x 4 | = | 16 | x 4 | = | **64** |
| $2^6$ | **x 6** | = | **64** | **x 6** | = | **384** |
| $2^8$ | x 8 | = | 256 | x 8 | = | 2048 |
| $2^{10}$ | x 10 | = | 1024 | x 10 | = | 10240 |
| $2^{12}$ | x 12 | = | 4096 | x 12 | = | 49152 |
| ... | | | | | | |

**Table B.1.** The three significant numbers for the genetic code: number 6 as the six-bit digital record of codons on the binary genetic code three (Rakočević, 1998); number 64 as the number of codons, and number 384 as the total number of bits on binary genetic code tree, and the total number of atoms within 20 amino acid molecules. (Note 1: First two columns – the realization of the principles of similarity and self-similarity.)

Note 2: (8–16 = - 8); (**64–64** = ±0); (384–256 = 2 x 64); (2048–1024 = 16 x 64); (10240–4096 = 96 x 64) …

**Appendix C**

In this appendix we add Table C.1 (in relation to Table C.2), which is in a way an extension of Shcherbak's Table of multiples of "Prime quantum 037" (which refers to the number of nucleons in the two classes of amino acids; four-codon and non-four-codon AAs) (Shcherbak, 1994, Table 1). We assume that the key features of Shcherbak's Table are the division by integer and the validity of the self-similarity principle, when it comes to the digits in number records ("037 versus 703" as a model). If we look at the first column in Shcherbak's Table (037, 370, 703) it is clear that the first two steps can be realized by all two-digit numbers while the third step (through module 9) is possible only for number 037; for example (037, 370, **703**) versus (038, 380, **722**).

As we can see, only the third step in Shcherbak's Table represents a unique opportunity, and here we show that this number (703) is at the same time the element of another arithmetic system (Table C.1, in relation to Table C.2). In this new system, in terms of division by integer, there appear only three numbers more: 105, 108 and 405; and when it comes to correspondence with the genetic code, the only significant number is 108 (see the section below the diagonal line in Table C.1).

---

$^{12}$ The fact that Nature "designed" everything in such a way that it equalized the number of atoms and bits (384), irresistibly leads us to Nicholas Negroponte (a Greek American architect, best known as the founder and Chairman Emeritus of Massachusetts Institute of Technology's Media Lab, and also known as the founder of the One Laptop Per Child Association, OLPC), who, 18 years ago came up with the idea of an inevitable relation between atoms and bits in the world (Negroponte, 1995). [Negroponte discusses the differences between bits and atoms. Atoms make up tangible physical objects and digital information, on the other hand, is made up of bits. He believes that all forms of information that are now made of atoms will eventually be made into bits.]

| | | | | | | | |
|---|---|---|---|---|---|---|---|
| 102\|201 | 103\|301 | 104\|401 | **105**\|501 | 106\|601 | 107\|701 | **108**\|801 | 109\|901 |
| | 203\|302 | 204\|402 | 205\|502 | 206\|602 | 207\|702 | 208\|802 | 209\|902 |
| 204−024 =180 | | 304\|403 | 305\|503 | 306\|603 | 307\|**703** | 308\|803 | 309\|903 |
| 204+**180** = 384 | | | **405**\|504 | 406\|604 | 407\|704 | 408\|804 | 409\|904 |
| 105 = 015 x 07 | | | | 506\|605 | 507\|705 | 508\|805 | 509\|905 |
| **108** = 018 x 06 | 037 x 01 = 037 | | | | 607\|706 | 608\|806 | 609\|906 |
| 405 = 045 x 09 | 037 x 10 = 370 | | | | | 708\|807 | 709\|907 |
| **703** = 037 x 19<br>407 = 037 x 11 | 037 x 19 = **703** | **108** x 64 = 3456+3456 | | | | | 809\|908 |
| [(64 x 6 = 384) (384 x 018 = 3456+3456)][$3^3+4^3+5^3 = 6^3$ (Plato's law)] | | | | | | | |

**Table C.1.** A possible natural numbers arrangement in relation to the "critical" Shcherbak's number 703 and its inversion 307. From only four division-integer cases, two (108 and 703) correspond to the number of atoms within constituents of the genetic code (384 in 20 amino acid molecules and 3456+3456 in 192+192 nucleotides). [Note 1: The significance of numbers 703 and 407 as it follows from Shcherbak's Table 1 in (Shcherbak, 1994), presented here as Table D.1. Note 2: Underlined patterns, first (102) in first row and second (204) in second row as half and whole number of atoms within 20 amino acid side chains, respectively.]

| | | | | | | | |
|---|---|---|---|---|---|---|---|
| 012\|210 | 013\|310 | 014\|410 | **015**\|510 | 016\|610 | 017\|710 | **018**\|810 | 019\|910 |
| | 023\|320 | 024\|420 | 025\|520 | 026\|620 | 027\|720 | 028\|820 | 029\|920 |
| | | 034\|430 | 035\|530 | 036\|630 | 037\|730 | 038\|830 | 039\|930 |
| | | | **045**\|540 | 046\|640 | 047\|740 | 048\|840 | 049\|940 |
| 510 = 015 x 34 | | | | 056\|650 | 057\|750 | 058\|850 | 059\|950 |
| **810 = 018** x 45 | **810** = 081 x 10 | | | | 067\|760 | 068\|860 | 069\|960 |
| 540 = 045 x 12<br>740 = 037 x 20 | **108** x 64 = 3456+3456 | | | | | 078\|870 | 079\|970 |
| Class II with 81 and Class I with 123 atoms (cf. Tab. 2.1) | | | | | | | 089\|980 |
| (A half set of AAs) 12**3** / **3**456 (A half set of nucleotides) | | | | | | | |

**Table C.2.** A possible natural numbers arrangement in relation to the Shcherbak's "Prime quantum 037" and its inversion 730. From only three division-integer cases, two (810 and 037) correspond to the number of atoms within constituents of the genetic code (81/123 and 3456/3456 in 20 amino acid molecules and in 192+192 nucleotides, respectively). [Note 1: The significance of number 037 and 740 as it follows from Shcherbak's Table 1 in (Shcherbak, 1994), presented here as Table D.1.]

**Appendix D**

In this appendix we add Table D.1 (Table 1 in: Shcherbak, 1994) and Table D.2 as the third possible pattern; all three patterns, corresponding with first three Shcherbak's numbers (first column in Table D.1: 037, 370, 703). The pattern ″703″ as in Table C.1; pattern ″037″ as in Table C.2, and pattern ″370″as in Table D.2.

| 1 | 2 | 3 | 4 | 5 | 6 | 7 | 8 | 9 |
|---|---|---|---|---|---|---|---|---|
| 037 | 074 | 111 | 148 | 185 | 222 | 259 | 296 | 333 |
| 10 | 11 | 12 | 13 | 14 | 15 | 16 | 17 | 18 |
| 370 | 407 | 444 | 481 | 518 | 555 | 592 | 629 | 666 |
| 19 | 20 | 21 | 22 | 23 | 24 | 25 | 26 | 27 |
| 703 | 740 | 777 | 814 | 851 | 888 | 925 | 962 | 999 |

**Table D.1.** The Shcherbak's Table of multiples of "Prime quantum 037" (Table 1 in: Shcherbak, 1994).

| 120\|021 | 130\|031 | 140\|041 | 150\|051 | 160\|061 | 170\|071 | 180\|081 | 190\|091 |
|---|---|---|---|---|---|---|---|
|  | 230\|032 | 240\|042 | 250\|052 | 260\|062 | 270\|072 | 280\|082 | 290\|092 |
|  |  | 340\|043 | 350\|053 | 360\|063 | **370**\|073 | 380\|083 | 390\|093 |
|  |  |  | 450\|054 | 460\|064 | 470\|**074** | 480\|084 | 490\|094 |
| 370 = 037 x 10 |  |  |  | 560\|065 | 570\|075 | 580\|085 | 590\|095 |
| 074 = 037 x 02 |  |  |  |  | 670\|076 | 680\|086 | 690\|096 |
|  |  |  |  |  |  | 780\|087 | 790\|097 |
| The numbers – multiples of 10 (and their inversions) |  |  |  |  |  |  | 890\|098 |

**Table D.2.** The third possible Table, corresponding with the Shcherbak's pattern ″370″.

**References**

Black, S.D., Mould, D.R., 1991. Development of hydrophobicity parameters to analyze proteins, which bear post- or cotranslational modifications. Anal. Biochemistry 193, 72-82.

Castro-Chavez, F., 2010. The rules of variation: Amino acid exchange according to the rotating circular genetic code. J. Theor. Biol. 264, 711-721.

Chechetkin, V.R., Lobzin, V.V., 2011. Stability of the genetic code and optimal parameters of amino acids. J. Theor. Biol. 269, 57-63.

Collins, F.S., 2006. The Language of God, Free Press, Bethesda, Maryland, USA.

Darwin, Ch., 1996. On the Origin of Species, Oxford UP.

Dragovich B, Dragovich A., 2010. p-Adic modeling of the genome and the genetic code. Comput. J. 53(4), 432-442; arXiv:0707.3043 [q-bio.OT].

Futuyma, D.J., 1979. Evolutionary Biology. 1st Ed. Sinauer Associates, Sunderland, Massachusetts.

Konopel'chenko, B.G., Rumer, Yu.B., 1975. Klassifikaciya kodonov v geneticheskom kode. Dokl. Akad. Nauk. SSSR 223, 471-474.

Kyte, J., Doollittle, R.F., 1982. A simple method for displaying the hydropathic character of a protein. J. Mol. Biol. 157, 105-132.

Marcus, S., 1989. Symmetry in the simplest case: the real line. Comp. Math. Applic. 17, 103-115.

Mišić, N. Ž., 2011. Nested numeric/geometric/arithmetic properties of shCherbak's prime quantum 037 as a base of (biological) coding/computing. Neuroquantology 9, 702-715.

Negadi, T., 2009. The genetic code degeneracy and the amino acids chemical composition are connected. Neuroquantology 7, 181-187; arXiv:0903.4131v1 [q-bio.OT].

Negroponte, N., 1995. Being Digital. Alfred A. Knopf, Inc., USA.

Popov, E.M., 1989. Strukturnaya organizaciya belkov. Nauka, Moscow (in Russian).

Pullen, S., 2005. Intelligent design or evolution? Free Press, Raleigh, N. Carolina, USA.

Rakočević, M.M., 1988. Three-dimensional model of the genetic code. Acta Biologiae et Medicine Experimentalis 13, 109-116.

Rakočević, M.M., 1997a. Two classes of the aminoacyl-tRNA synthetases in correspondence with the codon path cube. Bull. Math. Biol. 59, 645-648.

Rakočević, M. M., 1997b. Genetic code as a unique system. SKC, Niš (www.rakocevcode.rs)

Rakočević, M.M., 1998. The genetic code as a golden mean determined system. Biosystems 46, 283-291.

Rakočević, M.M., 2004. A harmonic structure of the genetic code. J. Theor. Biol. 229, 463-465.

Rakočević, M.M., 2011. Genetic code as a coherent system. Neuroquantology 9, 821-841. (www.rakocevcode.rs)

Rakočević, M.M., Jokić, A., 1996. Four stereochemical types of protein amino acids: synchronic determination with chemical characteristics, atom and nucleon number. J. Theor. Biol. 183, 345-349.

Shcherbak, V.I., 1994. Sixty-four triplets and 20 canonical amino acids of the genetic code: the arithmetical regularities. Part II. J. Theor. Biol. 166, 475-477.

shCherbak, V.I., Makukov, M A., 2013. The "Wow! signal" of the terrestrial genetic code. Icarus, 224, 228-242.

Swanson, R., 1984. A unifying concept for the amino acid code. Bull. Math. Biol. 46, 187-207.

Wetzel, R., 1995. Evolution of the aminoacyl-tRNA synthetases and the origin of the genetic code. J. Mol. Evol. 40, 545-550.

Woese, C.R. et al., 1966. On the Fundamental Nature and Evolution of the Genetic Code. Cold Spring Harb. Symp. Quant. Biol. 31, 723-736.